\documentclass[reprint,amsmath,amssymb, aps,prb]{revtex4-2}

\usepackage{xcolor}
\usepackage{hyperref}
\hypersetup{pdfborder=0 0 0,colorlinks=true,citecolor=blue,linkcolor=blue}

\usepackage{graphicx}
\usepackage{dcolumn}
\usepackage{bm}

\begin{document}

\title{Phase transitions of LaMnO$_3$ and SrRuO$_3$ from DFT + U based machine learning force fields simulations}

\author{Thies Jansen}\email{t.jansen@utwente.nl}
\author{Geert Brocks}
\author{Menno Bokdam}\email{m.bokdam@utwente.nl}
 \affiliation{University of Twente, Faculty of Science and Technology and MESA+ Institute 
for Nanotechnology, P.O. Box 217, 7500 AE Enschede, The Netherlands}

\date{\today}

\begin{abstract}
Perovskite oxides are known to exhibit many magnetic, electronic and structural phases as function of doping and temperature. These materials are theoretically frequently investigated by the DFT+U method, \textcolor{black}{typically in their ground state structure at $T=0$}. We show that by combining machine learning force fields (MLFFs) and DFT+U based molecular dynamics, it becomes possible to investigate the crystal structure of complex oxides as function of temperature and $U$. Here, we apply this method to the \textcolor{black}{magnetic transition metal compounds} LaMnO$_3$ and SrRuO$_3$. \textcolor{black}{We show that the structural phase transition from orthorhombic to cubic in LaMnO$_3$, which is accompanied by the suppression of a Jahn-Teller distortion, can be simulated with an appropriate choice of $U$. For SrRuO$_3$, we show that the sequence of orthorhombic to tetragonal to cubic crystal phase transitions can be described with great accuracy. We propose that the $U$ values that correctly capture the temperature-dependent structures of these complex oxides, can be identified by comparison of the MLFF simulated and experimentally determined structures.} 
 
\end{abstract}

\maketitle


\section{Introduction}
Complex transition metal oxides, in particular those with a perovskite structure, ($AB{\rm O}_3$), posses rich phase diagrams, in which many phases with  remarkable physical properties can be identified, such as high $T_c$ superconductivity \cite{Bednorz1986}, ferromagnetism \cite{Wang2015}, ferroelectricity, or multiferroic properties, and metal to insulator transitions under external fields, pressure, doping or temperature change  \cite{Urushibara1995}. Many such properties are not only interesting because of fundamental science, but they may also lead to new types of applications in oxide electronics \cite{Lorenz2016,Coll2019}. The properties arise due to the intricate interplay between the structure and  the electronic interactions, leading to charge order, magnetic order, and/or orbital order  \cite{Hwang2012, Coll2019}. The emergence of different structural phases is therefore strongly tied to the nature of the electronic structure. 

Many of the interesting phases occur at a finite temperature ($T>0$ K), and over a limited temperature range. This calls for simulations that can describe the physics at elevated temperatures. Ordinary molecular dynamics (MD) is of limited use, as it is based upon fixed force fields, which do not consider changes in the electronic structure. In the materials discussed above, the latter is intimately coupled to, if not driving, the phase transitions.  \emph{Ab initio} MD would be an option, but unfortunately it is very time consuming, as in every MD step a density functional theory (DFT) calculation has to be performed to obtain the total energy and the forces on the atoms. In practice, only processes occurring over very small time scales can then be simulated. 

Recent developments in machine-learning (ML) inter-atomic potentials open up new computational routes \cite{Behler:prl07,Bartok:prl10,Rupp:prl12,Bartok:prb13}. Here we apply an approach where a ML model of the potential energy surface is trained on-the-fly during \emph{ab initio} MD runs \cite{Li:prl15} using limited supercell sizes and time scales. Subsequently, the machine learned force field (MLFF) is then used to perform simulations on much larger supercells and time scales. On the one hand, this approach takes the advantages of \emph{ab initio} MD, i.e., it incorporates the information about the electronic structure. On the other hand, a MLFF greatly reduces the computational resources required to run the MD simulations, thereby enabling the ensemble sizes and time scales required to draw statistically valid conclusions. 

In the past three years this type of on-the-fly MLFFs has been successfully employed in various applications, such as phase transitions in hybrid lead halide perovskites \cite{Jinnouchi:prl19,Bokdam:jpcc21}, superionic transport in AgI\cite{Vandermause:npjcm20}, melting of solids \cite{Jinnouchi:prb19}, surface reconstruction of palladium adsorbed on silver \cite{Lim:jacs20}, NMR $^1$H-$^1$H dipolar coupling in mixed hybrid lead halide perovskites \cite{Grueninger:jpcc21}, the structure-band gap relationship in SrZrS$_3$ \cite{Jaykhedkar:jmcc22}, and the catalysis dynamics of H/Pt \cite{Vandermause:natc22}.

\textcolor{black}{The MLFFs in those studies are based upon results obtained with standard generalized gradient approximation (GGA) functionals. Such functionals, however, tend not to describe interactions between localized electrons very well, such as they occur on transition metal ions in oxides, for instance. The description can be improved significantly by adding a parametrized model electron-electron interaction on the metal ions, leading to the DFT+U method \cite{Anisimov1997}. The latter is designed to capture some of the essential physics of electrons localized in atomic shells \cite{Himmetoglu2013, Anisimov1991_1, Wenging2013, Anisimov1997, Solovyev1994}. Although, in principle, more accurate methods for describing local correlations exist \cite{Wenging2013, Mandal2019}, the DFT+U method is cost efficient and currently the practical electronic structure method of choice for calculations on systems with sizeable unit cells involving localized electrons, as they occur in many transition metal compounds \cite{KirchnerHall2021, Wenging2013, Mandal2019}.}

\textcolor{black}{In this paper we address the question whether the on-the-fly MLFF method can learn a representation of the potential energy surface of materials studied by DFT+U, and correctly capture the physics of the interplay between the electronic and crystal structure at finite temperatures. Several first-principles techniques exist that allow for a calculation of the on-site Coulomb or exchange parameters ($U,J$), based upon linear response \cite{Cococcioni2005},
unrestricted Hartree-Fock \cite{Mosey2007, Mosey2008}, constrained random-phase approximation \cite{Aryasetiawan2006, Miyake2008, Esroy2011}, or machine learning \cite{Yu2020}, but in practice these parameters are often treated as empirical. Furthermore, there may be a spread in the parameter values obtained with different theoretical approaches, and not all values are satisfactory \cite{Shi2017, Selcuk2015, Shi2015}. Finite temperature simulations using MLFFs might then allow for testing those values, while producing data that can be compared to experiment \cite{Artrith2022}.}

We study two archetypal complex transition metal oxide perovskites, LaMnO$_3$ and SrRuO$_3$, for which \textcolor{black}{one expects to see a varying degree of electron localization on the transition metal ion. Both these materials show structural phase transitions as a function of temperature, and, as we will see, the proper simulation of those requires an adequate description of that localization.}  

The transition metal Mn in LaMnO$_3$ has a very localized 3d shell with a $d^4$ configuration in a high spin state (three electrons in $t_{2g}$ states, and one in a $e_g$ state, all parallel, leading to a maximum magnetic moment $M=4$ $\mu_\mathrm{B}$). Bulk LaMnO$_3$ is an insulator with antiferromagnetic order at low temperature, and a Neél temperature of 140~K \cite{Wollan1955}. Because the $e_g$ states are filled with a single electron, the MnO$_6$ octahedra in LaMnO$_3$ are Jahn-Teller (JT) distorted \cite{Rodriguez1998}. 

It results in orbital ordering, which controls the orthorhombic structure observed for LaMnO$_3$ at temperatures $T < 750$ K. LaMnO$_3$ shows a somewhat unusual structural phase transition at $T = 750$ K, where the JT distortion is suppressed and the lattice parameters become equal. However, in this 'metric cubic' phase octahedral tilting remains present, and the symmetry of the crystal remains orthorhombic \cite{Rodriguez1998}. Using results from our MLFF MD simulations, we present a direct comparison to crystal structure data obtained from X-ray diffraction experiments at finite temperatures.
 
The 4d transition metal Ru in the perovskite SrRuO$_3$ also has a $d^4$ electron configuration. The Ru 4d shell is much less localized than the Mn 3d shell, and the properties of SrRuO$_3$ are markedly different from those of LaMnO$_3$.  The Ru 4d electrons adopt a low spin configuration with all electrons in $t_{2g}$ states. This can lead to a maximum atomic magnetic moment $M=2$ $\mu_\mathrm{B}$, which in SrRuO$_3$ decreases somewhat, because of hybridization of atomic states and delocalization of electrons \cite{Kennedy1998}. 

SrRuO$_3$ is a half-metallic ferromagnet \cite{Koster2012, Allen1995, Kanbayasi1976, Gan1998, Takiguchi2020, Rondinelli2008}. It exhibits an orthorhombic phase at lower temperatures \cite{Jones1989}, becomes tetragonal at 820~K and cubic at 950~K \cite{Kennedy1998}. In contrast to the phase transition in LaMnO$_3$, the sequence of structures characterizing the phase transitions in SrRuO$_3$ is more standard for perovskite structures.  The low temperature, orthorhombic, phase is dictated by structural parameters (Goldschmidt tolerance factors) that trigger octahedral tilting. Each phase transition removes one or more of these tilts, which increases the symmetry of the crystal. 

In this work we show that, for these two, electronically very different,  complex oxides, MLFFs can capture the physics of the structural phase transitions. The simulated phase transition temperatures depend on the $U$ values chosen, demonstrating on the one hand the strong connection between structure and electron \textcolor{black}{localization}, and, on the other hand, providing the possibility of assessing the \textcolor{black}{value of $U$} by means of the transition temperatures.

\section{Methods}
Our MD simulations are divided into two stages, i.e. a first, training stage, and a second, production stage. The first stage aims at the on-the-fly training of the MLFF, using the techniques described in Refs. \cite{Jinnouchi:prl19,Jinnouchi:prb19} with \textcolor{black}{parameter settings motivated by those studies, such as the use of} a $2\times2\times2$ LaMnO$_3$ or SrRuO$_3$ supercell. The MLFF potential is constructed based on a variant of the GAP-SOAP method \cite{Bartok:prl10,Bartok:prb13} with a Gaussian width of 0.5~\r{A}. The local atomic configurations are described within a cutoff radius of 6 and 5~\r{A} for the two- and three-body terms, respectively. To study the influence of the $U$ value, a MLFF is generated for each of the values mentioned in the result section. The regression results of all the learned models can be found in the Supplementary Material Sec. S1.

Collinear spin-polarized DFT calculations are carried out using \textsc{vasp~6.3} \cite{Kresse1993, Kresse1996}, which applies the projector augmented-wave method \cite{Blochl:prb94b} to calculate the electronic states. A $500$~eV cut-off energy for the plane-wave basis, and a gaussian smearing of $\sigma=0.05$~eV are used. A $2 \times 2 \times 2$  and a $4 \times 4 \times 4$ $\Gamma$-centered $k$-point grid are used for LaMnO$_3$ and SrRuO$_3$, respectively. \textcolor{black}{The density of states (DOS) of both compounds are calculated with a fully relaxed primitive orthorhombic unit cell and a $4 \times 4 \times 4$ $\Gamma$-centered $k$-point grid.}

\textcolor{black}{To construct the MLFFs as described in \cite{Jinnouchi:prl19} an explicit electronic structure calculation, in this case} DFT+U, is done whenever the Bayesian uncertainty in the predicted forces by the MLFF us too large. The Perdew-Burke-Ernzerhof (PBE) generalized gradient approximation (GGA) is used for LaMnO$_3$ \textcolor{black}{as part of the DFT+U description}, and PBEsol for SrRuO$_3$. \textcolor{black}{Generally, PBEsol tends to give somewhat better lattice parameters than PBE, but we mainly based our choice of functionals on previous studies on these materials, see the next paragraph.}

\textcolor{black}{For LaMnO$_3$ a Hubbard term according to the rotationally invariant description of Dudarev \textit{et al.}  \cite{Dudarev1998} is used, with $U_{\rm eff}=U-J=3.5$~eV for the Mn~$3d$ shell, which is a choice guided by previous work \cite{Chen2017}.} An A-type antiferromagnetic configuration is used for the magnetic ordering in LaMnO$_3$ at $T=0$. \textcolor{black}{For SrRuO$_3$ a Hubbard term is introduced according to the DFT+U method of Liechtenstein \textit{et al.} \cite{Liectenstein1995}, with parameters $U$ and $J$ for the Ru $4d$ orbitals set to $2$ eV and $0.6$ eV, respectively, again a choice that is consistent with previous studies \cite{Takiguchi2020, Solovyev1994, Rondinelli2008, Granas2014}}. Here, a ferromagnetic ordering is set for $T=0$. 

In the second stage, after the MLFFs have acquired sufficient accuracy, production runs are made using $6\times6\times6$ supercells, where the forces are calculated solely with the MLFFs, without resorting to \textcolor{black}{DFT+U} calculations. The isothermal-isobaric MD simulations with timesteps of 0.5~fs are performed with a Langevin thermostat, and in calculating the phase transitions, the temperature is varied from 100~K to 1100~K at a rate of 0.5~K/ps under constant atmospheric pressure.

\begin{figure}
    \centering
    \includegraphics[scale = 0.85]{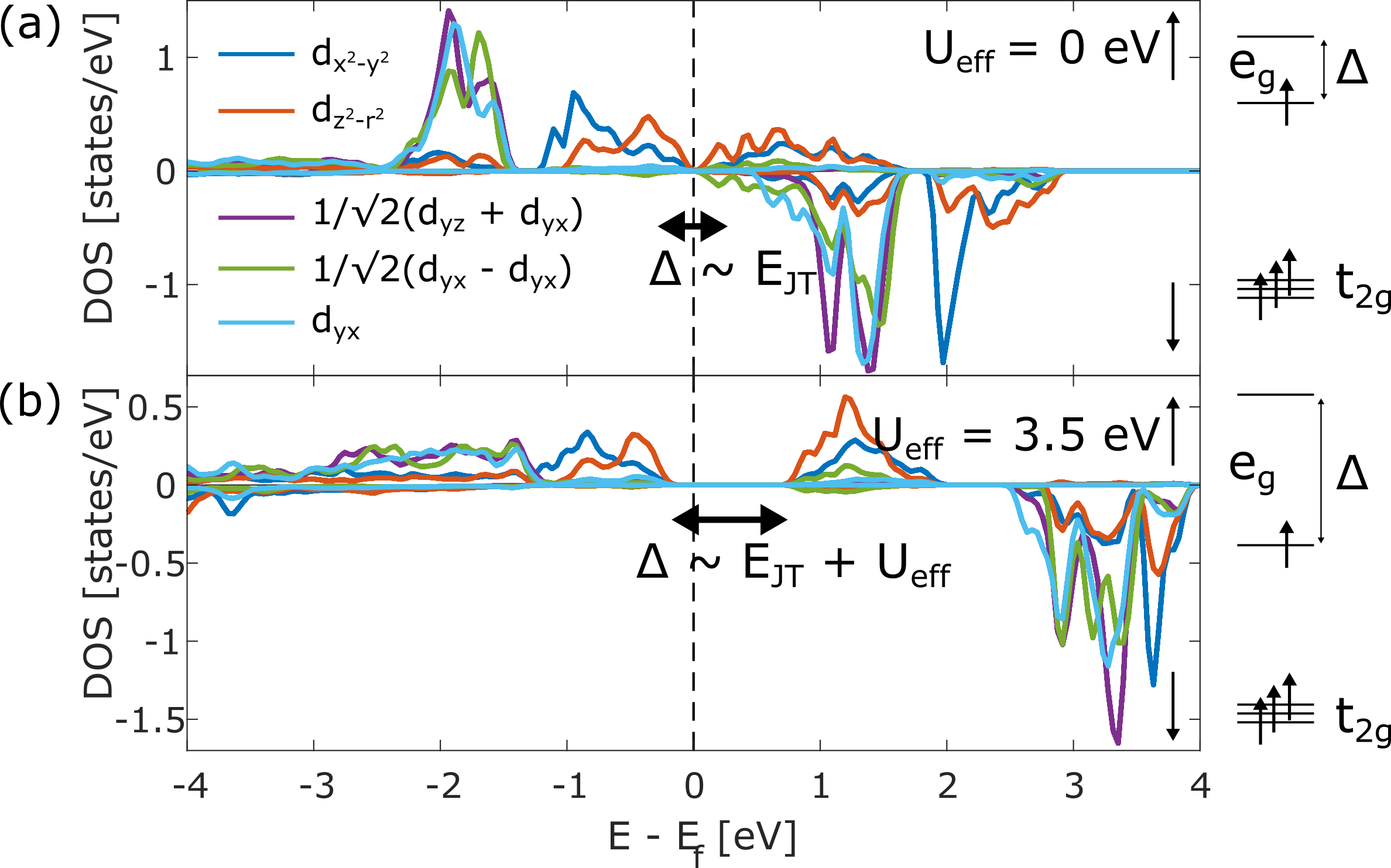}
    \caption{Density of states (DOS) of LaMnO$3$ \textcolor{black}{in its ground state structure},  projected on the $d$ orbitals of \textcolor{black}{a single Mn atom} for (top) $U_\mathrm{eff} = U-J= 0$ eV and (bottom) $U_\mathrm{eff} = 3.5$ eV. \textcolor{black}{The right-hand side shows schematically the splitting of the $d$ orbitals due to the crystal field, including the effect of the Jahn-Teller (JT) distortion.}}
    \label{fig:LMOdos}
\end{figure}

\begin{figure*}[t]
    \centering
    \includegraphics[scale = 0.9]{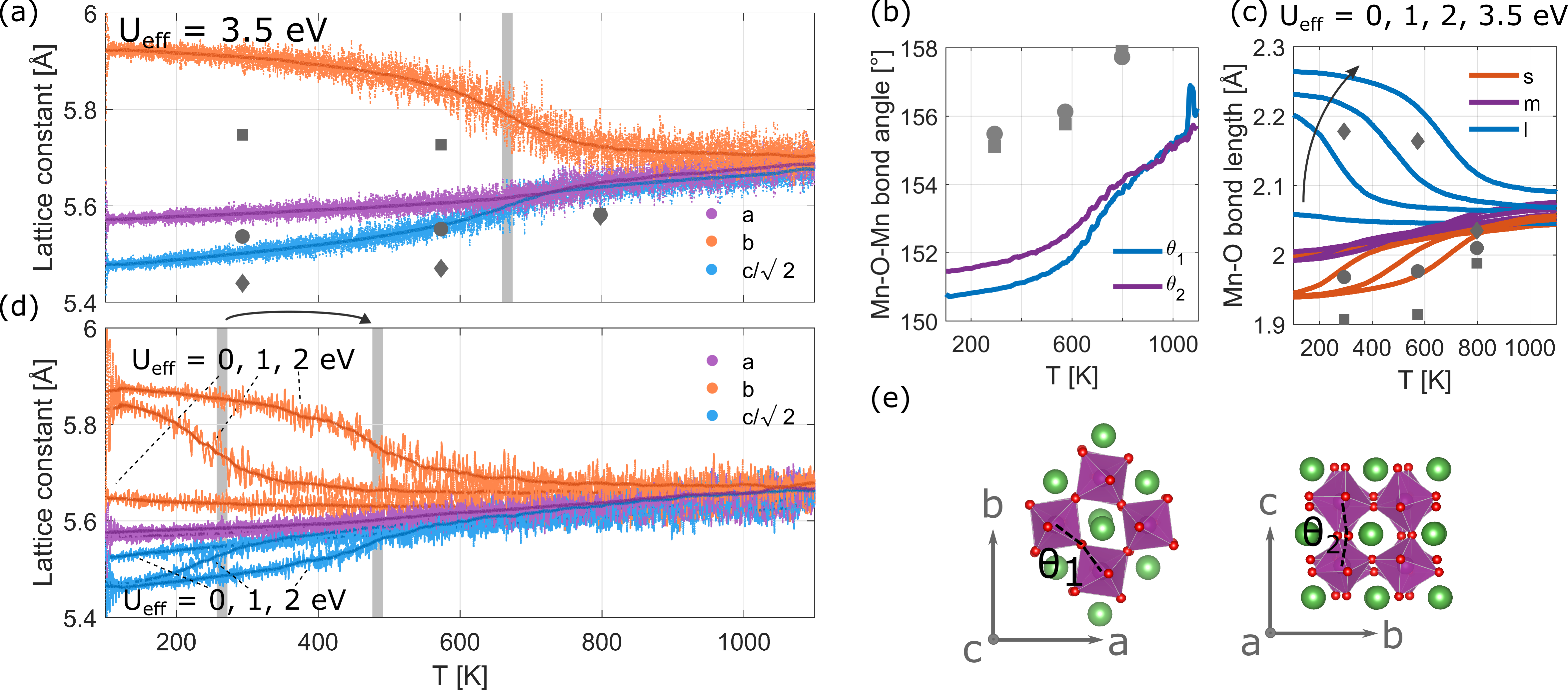}
    \caption{Results of MD simulations on LaMnO$_3$, using machine learning force fields. (a) Lattice parameters as function of temperature, \textcolor{black}{sampled} at every tenth MD step (5 fs). The solid line is a moving average of 500 steps. The symbols $\bullet$, $\blacksquare$,  and $\blacklozenge$ indicate the experimental $a$, $b$ and $c$ lattice parameters respectively, taken from \textcolor{black}{Ref.} \cite{Rodriguez1998}. (b) Moving average of bond angles as function of temperature, the symbols indicate the experimental data. (c) Moving average of bond length as function of temperature for $U_\mathrm{eff} = 1, 2, 3,$ and $3.5$ eV. The symbols indicate the experimental data. (d) Lattice parameters as function of temperature for $U_\mathrm{eff} = 0, 1$ and $2$ eV. (e) \textcolor{black}{Definitions of the bond angles plotted in (b).}}
    \label{fig:1}
\end{figure*}

\section{Results}

\subsection{LaMnO$_3$}
The basic electronic structure of LaMnO$_3$ \textcolor{black}{in its ground state structure}, resulting from a DFT+U calculation, is shown in Fig. \ref{fig:LMOdos}. The octahedral coordination of Mn gives the standard $t_{2g}$-$e_g$ split d orbitals, where for the Mn d$^4$ configuration the lower $t_{2g}$ states are half filled, and the upper $e_g$ states contain one electron, resulting in a high spin state, with a magnetic moment $M=4$ $\mu_B$ per Mn atom, and anti-ferromagnetic ordering in the ground state of moments on different Mn atoms.

The singly occupied $e_g$ states give rise to a Jahn-Teller (JT) distortion and concomitant orbital ordering, which has a marked influence on the density of states (DOS),  Fig. \ref{fig:LMOdos}. In a calculation without Hubbard terms, $U_\mathrm{eff} = 0$ eV, the JT distortion is small even at $T=0$, giving a difference in Mn-O bond lengths of $\sim 0.07$ \AA\ only. This results in a small splitting of the $e_g$ orbitals, and a corresponding vanishing band gap $\Delta$, whereas the experimental band gap is 1.2 eV \cite{Tokura2001}.  

Setting $U_\mathrm{eff} = 3.5$ eV increases the JT distortion significantly, enlarging the difference in Mn-O bond lengths to $\sim 0.3$ \AA. Correspondingly, the DFT+U band gap increases to 0.88 eV, which is much closer to the experimental value. At $T=0$, these calculations give an orthorhombic structure for LaMnO$_3$, with optimized lattice constants $a=5.57$, $b=5.90$, and $c=7.73$ \AA. The experimental low temperature structure is indeed orthorhombic, with lattice parameters at $T=300$ K of $a=5.54$, $b=5.74$, and $c=7.69$ \AA\ \cite{Rodriguez1998}. The overestimation of the lattice parameters by the DFT calculation is a common consequence of the PBE functional.

\textcolor{black}{The size of the band gap is coupled to that of the JT distortion, and both are affected by the effective localization of the electrons in the Mn d orbitals, as controlled by the value of $U$. It should be noted, however, that when directly comparing with experimental data, calculating band gaps with small unit cells may give slightly misleading results, and supercells that have a substantial number of local structural and magnetic degrees of freedom may be required \cite{Oleksandr2023, Varignon2019}.}

A MD simulation using the MLFF, while increasing the temperature linearly, allows us to study the structural transition of LaMnO$_3$. Figure \ref{fig:1}(a) shows the lattice parameters of LaMnO$_3$, as function of temperature. How the orthorhombic cell parameters are extracted from the $6\times 6\times 6$ supercell simulations is explained in the Supplementary Material, Sec. S2.  At low temperature, the MD simulations clearly maintain an orthorhombic structure for LaMnO$_3$, which is in agreement with experiment. Every tenth MD step is plotted in Fig.~\ref{fig:1}(a) and a moving average of 500 steps is used to show the trend. In the entire temperature range the lattice vectors remain perpendicular to each other. 

The MD calculations preserve the orthorhombic (Ort) phase in the temperature range $T\lesssim 700$ K, and give a transition to a metric cubic phase (Cub) above 700 K, \textcolor{black}{where} the lattice parameters become equal, but the crystal structure does not adopt a cubic symmetry.  Experimentally, a Ort-Cub phase transition is observed at $T=750$ K \cite{Rodriguez1998}. In the MD run, an extremely sharp transition is not expected, due to the still relatively fast heating rate that is demanded for in the simulations \cite{Jinnouchi:prl19,fransson2023phase}. The temperature at which the change of the lattice parameters as function of temperature is at a maximum (indicated by the grey bar at $T\approx 680$ K in Fig. \ref{fig:1}(a)) is used to pinpoint the Ort-Cub phase transition temperature. 

From the perspective of the lattice parameters, the global structure becomes cubic for temperatures above the transition temperature. However, the microscopic structure reveals that the symmetry of the crystal is still orthorhombic in the Cub phase, which is in agreement with experiment \cite{Rodriguez1998}. In the low temperature Ort phase the MnO$_6$ octahedra are significantly distorted and tilted, with all Mn-O-Mn bonds between neighboring octahedra close to $150^\mathrm{o}$ (or $360^\mathrm{o}-150^\mathrm{o}$, Fig. \ref{fig:1}(e)). In the high temperature Cub phase, the tilting persists, with angles close to $155^\mathrm{o}$, Fig \ref{fig:1}(b). For a description of the averaging procedure used to obtain the tilting angles see the Supplementary Material, Sec. S3. At higher temperatures, the distribution of angles around $155^\mathrm{o}$ broadens. In addition, a secondary distribution appears at $(360 -155)^\mathrm{o}$, which is an indication of octahedra flipping over their tilt, see the Supplementary Material, Sec. S3.  

The microscopic structural changes in the Orb-Cub phase transition can also be monitored by inspecting the Mn-O bond lengths during the MD run, see Fig. \ref{fig:1}(c). The calculations predict a significant JT distortion for lower temperatures in all MnO$_6$ octahedra, where Mn-O bond lengths can be divided into three classes, i.e., a long ($\sim 2.27$ \AA), medium ($\sim 2.00$ \AA) and short ($\sim 1.94$ \AA) bond. This JT distortion is almost completely suppressed in each octahedron for $T \gtrsim 800$~K, in agreement with the experiment. Nevertheless, as described above, the tilting of octahedra persists even at higher temperatures.

The structural phase transition can be linked to the the influence of electron \textcolor{black}{localization, by evaluating this transition for different values of $U_\mathrm{eff}$}. Figure \ref{fig:1}(d) shows the lattice parameters as function of temperature, characterizing the Ort-Cub phase transition, as obtained from MD runs with MLFFs, trained for DFT calculations with different setting of the Hubbard parameter, $U_\mathrm{eff} = 0,1$ and $2$ eV, respectively. Ignoring the on-site Hubbard term, $U_\mathrm{eff} = 0$, the orthorhombic distortion is severely underestimated already at a very low temperature, and there is no clear transition to a different phase at any temperature. 

If one increases the on-site repulsion $U_\mathrm{eff}$ to $1$ and $2$~eV, the orthorhombic distortion at low temperature becomes much more pronounced. Moreover, as indicated by the grey bars in Fig. \ref{fig:1}(d), the transition temperature for the Ort-Cub phase transition increases with increasing $U_\mathrm{eff}$, from 270 K, 490 K, to 680 K for $U_\mathrm{eff}$ to $1$, $2$ and $3.5$~eV, respectively. Comparing this to experiment \cite{Rodriguez1998}, the latter value thus seems the most reasonable. Figure \ref{fig:1}(c) shows that the size of the JT distortion at  low temperature increases monotonically with increasing $U_\mathrm{eff}$, and that the Ort-Cub transition at higher temperature is accompanied by a suppression of the JT distortion for all values of $U_\mathrm{eff}$.  

\subsection{SrRuO$_3$}
\begin{figure}
    \centering
    \includegraphics[scale = 1]{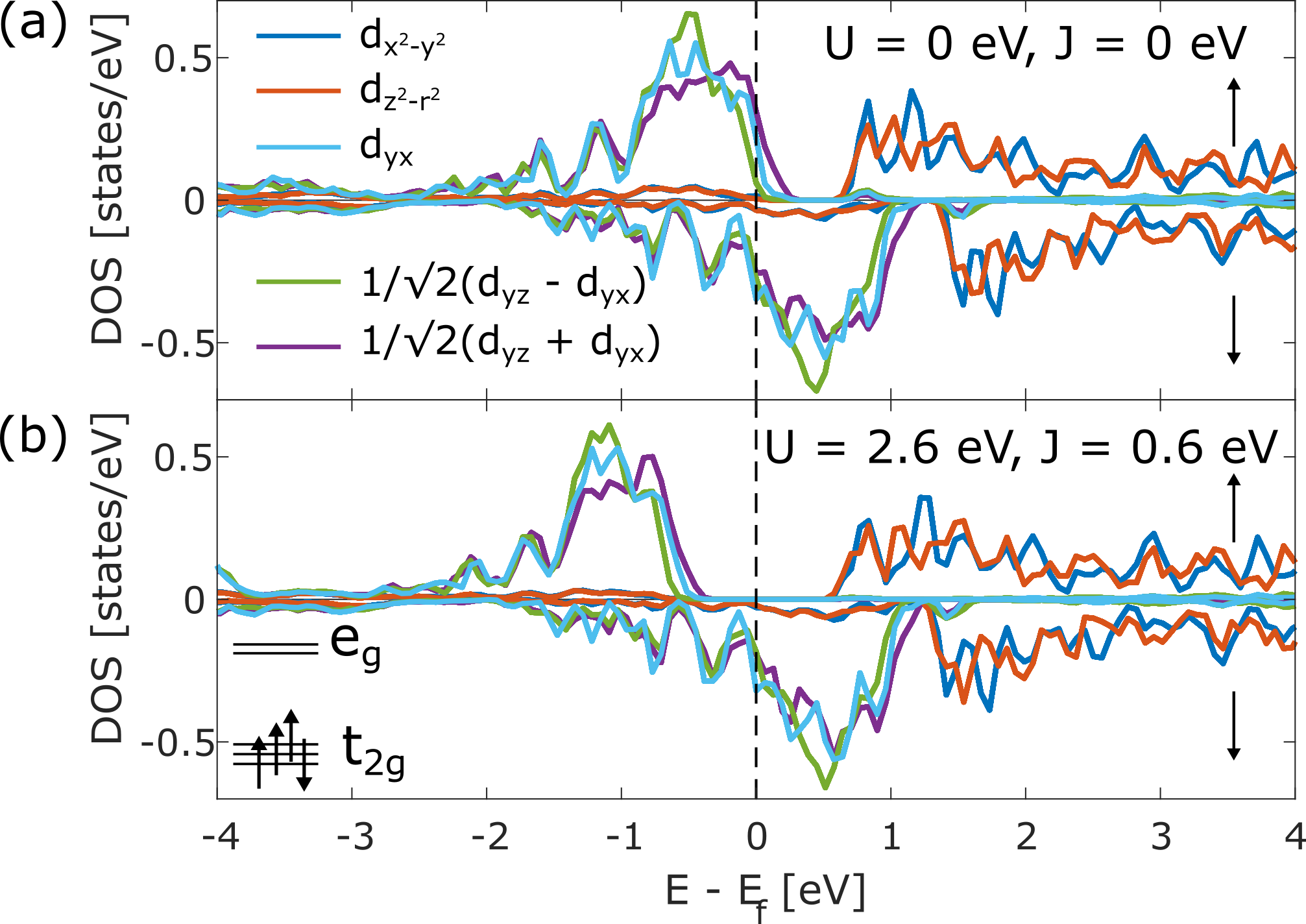}
    \caption{Density of states (DOS) of SrRuO$_3$ projected on the individual $d$ orbitals of Ru for $U = J = 0$ eV and $U = 3.5$ and $J = 0.6$ eV.}
    \label{fig:SROdos}
\end{figure}

Although in both cases the transition metal ion has d$^4$ configuration, the basic electronic structure of SrRuO$_3$ differs markedly from that of LaMnO$_3$, compare Figs. \ref{fig:SROdos} and \ref{fig:LMOdos}. Setting the Coulomb and exchange parameters $U,J = 2.6,0.6$ eV, as in previous calculations \cite{Takiguchi2020, Solovyev1994, Rondinelli2008, Granas2014}, the Ru ions adopt a low spin state, with the four d electrons occupying $t_{2g}$ states. \textcolor{black}{The calculated magnetic moments on the Ru atoms are 1.44 $\mu_B$}. There is no occupation of the $e_g$ orbitals, and no JT distortion. SrRuO$_3$ is a half-metallic ferromagnet in its ground state, see Fig. \ref{fig:SROdos}(b).

In contrast, neglecting the onsite terms, $U,J = 0,0$ eV, in the calculations, SrRuO$_3$ becomes a standard metallic ferromagnet. The $e_g$ states remain unoccupied, but in addition the majority spin $t_{2g}$ states are no longer fully occupied, see Fig. \ref{fig:SROdos}(a). Correspondingly, the calculated magnetic moments \textcolor{black}{(1.30 $\mu_B$)} are smaller than in the case discussed in the previous paragraph.

Like LaMnO$_3$, SrRuO$_3$ shows structural phase transitions as a function of temperature, although quantitatively the structural changes are smaller than those in LaMnO$_3$. The low temperature structure of SrRuO$_3$ is characterized by a relative tilting of RuO$_6$ octahedra, mainly driven by geometric considerations (Goldschmidt tolerance factors). Experimentally, for $T < 600$ K, SrRuO$_3$ adopts an orthorhombic phase.  In the temperature range $600$ K $ < T < 900$ K, the structure is tetragonal, and for $T > 900$ K the structure becomes cubic \cite{Jones1989,Kennedy1998}. 

This sequence of phases, which is not uncommon in perovskites, is determined by a gradual removal of octahedral tilting. The orthorhombic phase is characterized by two tilt angles, and is labeled $b^-b^-a^+$ in Glazer notation \cite{Kennedy1998}. The tetragonal phase is a structure with one tilt angle, labeled $a^0a^0c^-$, whereas all tilting disappears in the cubic structure.  For the Pb-halide based perovskites, similar subtle structural changes and the related phase transitions have been shown to be well described by MLFFs \cite{Bokdam:jpcc21}. The question here is whether \textcolor{black}{the $U,J$ terms} play a role in these phase transitions and give the right transition temperatures. 

\begin{figure}
    \centering
    \includegraphics[scale = 0.85]{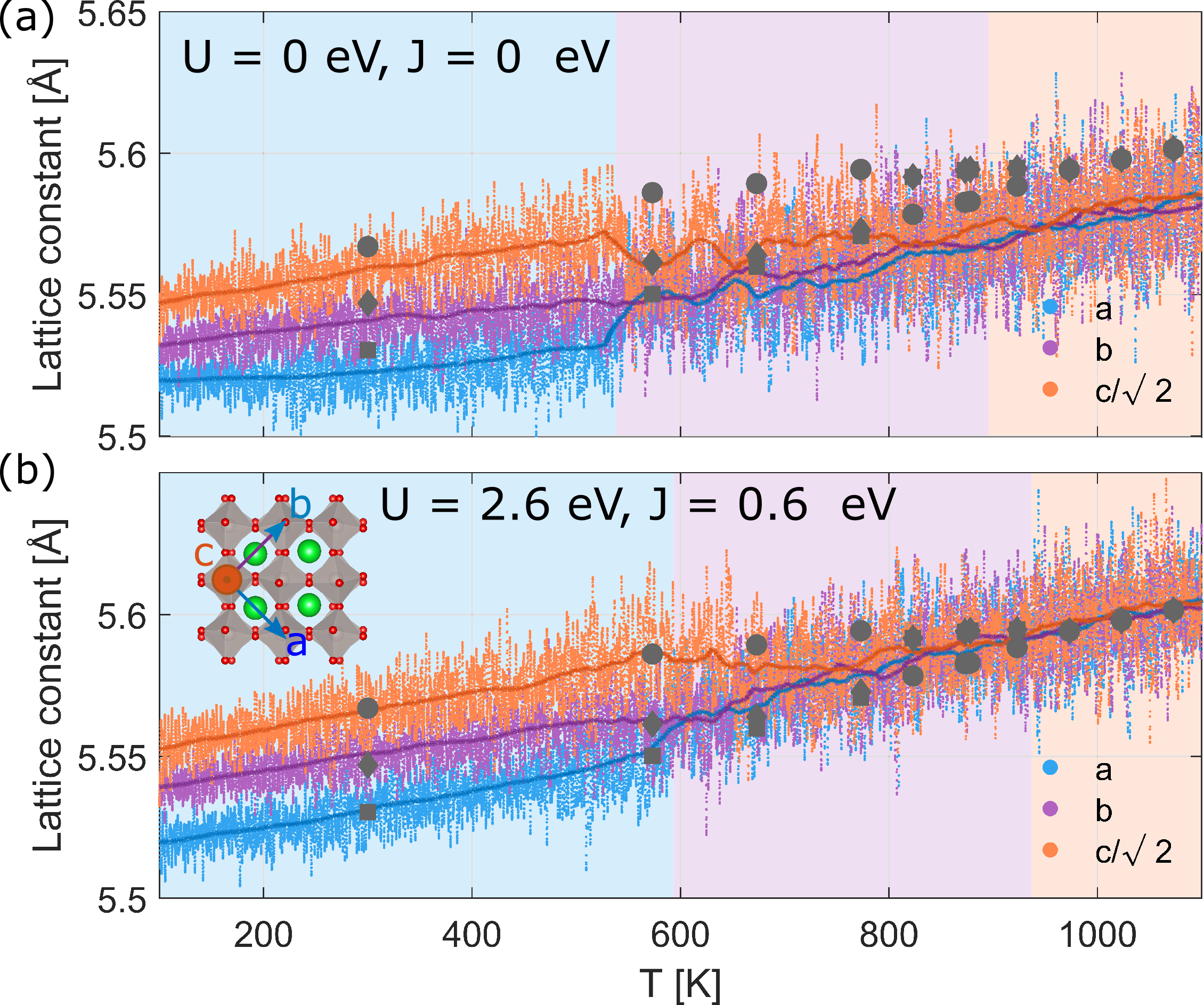}
    \caption{Results of MD calculations on SrRuO$_3$, using machine learning force fields, for the cases $U = 2.6$, $J = 0.6$ eV and $U = J = 0$ eV. Lattice constants as function of temperature. The solid lines are a moving average of 500 MD steps. The symbols $\bullet$, $\blacksquare$,  and $\blacklozenge$ indicate the experimental $a$, $b$ and $c$ lattice parameters respectively, taken from \cite{Kennedy1998, Jones1989}. }
    \label{fig:SRO}
\end{figure}

Figure \ref{fig:SRO} shows the evolution of the lattice constants of SrRuO$_3$ as function of temperature for the cases $U,J = 0,0$ and $U,J = 2.6,0.6$~eV, as calculated with MLFFs, using the same procedure as for LaMnO$_3$. For the case of $U,J = 2.6,0.6$~eV the calculations show a clear orthorhombic phase for $T < 600$~K, a clear cubic phase is present for $T > 900$~K, and for $600$~K $ < T < 900$~K the structure is tetragonal, in agreement with experiment. Moreover, also quantitatively the lattice parameters are in good agreement with experimental results \cite{Jones1989,Kennedy1998}. 

For $U,J = 0,0$ eV the agreement with experiment is less gratifying. The calculated lattice parameters are too small over the whole temperature range, and the temperatures at which the phase transitions occur, are underestimated by $\sim 100$ K. The effect of \textcolor{black}{including $U,J$ terms} on the finite temperature behavior of SrRuO$_3$ is less dramatic than in the case of LaMnO$_3$. Nevertheless also in SrRuO$_3$, including on-site electron-electron interactions is important to obtain quantitative agreement with experiment.

\section{Discussion}
The MD simulation with the MLFF predicts the sequence of structural phases of LaMnO$_3$ as a function of temperature in good agreement with experiment, provided the MLFF is trained on-the-fly in an \emph{ab initio} MD simulation based upon a DFT+U description of the electronic structure. The on-site Coulomb interaction $U$ on the Mn ions plays an essential role in \textcolor{black}{improving  the description of the finite temperature structures over those given by regular DFT functionals (such as PBE)}.
 
The low temperature phase is dominated by a JT distortion of the MnO$_6$ octahedra, which dictates orbital ordering resulting in an orthorhombic structure. \textcolor{black}{Using only the PBE functional}, the ground state of LaMnO$_3$ is still described correctly as an anti-ferromagnetic insulator. However, the JT distortion is severely underestimated, as are the band gap, and the phase transition temperature. 

Upon introducing $U$, the JT distortion is enlarged to a quantitatively realistic value. One of the consequences of the on-site terms in the DFT+U formalism is to more localize the d-electrons on the Mn ions in comparison to a standard DFT GGA functional, thereby amplifying the effects of such a localization, such as the size of the JT distortion of the MnO$_6$ octahedra, and its consequences for the finite temperature behavior of LaMnO$_3$ as discussed above. 

The size of the effects scale with the size of $U$. For $U=3.5$ eV, the structure of the initial orthorhombic phase and the transition to the metric cubic phase around 750~K are \textcolor{black}{reasonably well described}. This structural transition is accompanied by a suppression of the JT distortion on each MnO$_6$ octahedron. We conclude that the JT distortion is the driving force behind the dissimilar lattice parameters at lower temperatures. 

The temperature at which the JT is suppressed, and thus the lattice parameters become equal, is proportional the electronic energy gained by breaking the symmetry of the lattice in the form of a JT distortion. This energy becomes larger with increasing $U$ \cite{Weng2015, Dagotto2001, Hotta2006}. So, it is reasonable then that, if $U$ is made smaller, the JT distortion, the electronic band gap, and the Ort-Cub transition temperature, all become smaller. The value $U=3.5$ eV gives a transition temperature that is quite close to the experimental value.

\textcolor{black}{We have used the simplest formulation of the DFT+U method in the calculations on LaMnO$_3$ \cite{Dudarev1998}, and, although the structural phase transition temperature is well predicted, there may be room for improvement, such as orbital dependent Coulomb $U$ parameters, the introduction of exchange $J$ parameters, or by also applying $U$ terms to the ligand orbitals \cite{Mellan2015, Linscott2018, Orhan2020}. Other possibilities can be found in the use of spatially dependent values of U \cite{Heather2011, Kulik2015} or in the introduction of intersite terms \cite{Heather2011_2}.} 

Also for SrRuO$_3$ the MD simulation with the MLFF predicts the sequence of structural phases well as function of temperature. The number of d electrons on the Ru ions is formally the same as on the Mn ions in LaMnO$_3$, but the effects of \textcolor{black}{on-site $U$ and $J$ terms} are much smaller. There is no JT distortion of the RuO$_6$ octahedra, and SrRuO$_3$ is a (ferromagnetic) metal. SrRuO$_3$ has an orthorhombic structure at low temperature, a cubic structure at high temperature, and a tetragonal structure at intermediate temperatures. As in many perovskites, this sequence of phases is dictated by geometric structural parameters. 

The transitions between these phases are correctly described by the MLFF. Using the parameters $U,J=2.6,0.6$ eV gives excellent agreement with experiment of the lattice parameters as a function of temperature. In contrast, using $U,J=0,0$ eV, the agreement becomes much less impressive, and, in particular, the temperatures at which the phase transitions are predicted to occur, are too low. We conclude therefore, that also in the case of SrRuO$_3$ it is important to include \textcolor{black}{explicit on-site electron terms} to correctly describe the finite temperature behavior of the material. 

In addition, a clear difference in the $T = 0$ K band structure is observed between the DFT and DFT+U method. DFT predicts SrRuO$_3$ to be a conventional metallic ferromagnet. In contrast, the DFT+U method states that SrRuO$_3$ is a half-metallic ferromagnet, as reported earlier by other calculations \cite{Tay2006, Takiguchi2020} and suggested by experiments \cite{Nadgorny2003, Raychaudhuri2003, Sanders2005}. 

\textcolor{black}{In the calculations discussed in this paper, we have used relatively simple DFT+U methods, which of course have their limitations \cite{Pavarini2012, Hong2015, Kulik2015, Himmetoglu2013, Ryee2018}. In particular, one could mention the use of a single $U$ parameter \cite{Kasinathan2007, Ylvisaker2009, Bousquet2010, Tompsett2012}, the PAW projector functions used to define the on-site terms \cite{Wang2016, Pickett1998, Tablero2008, Han2006, Regan2010}, and treating the Coulomb and exchange $U$ and $J$ as adaptable parameters \cite{Wang2016, Silva2007,Stahl2020, Meng2016}. In principle, such parameters can be calculated from first principles \cite{Pickett1998, Cococcioni2005, Vaugier2012}. One may then consider finite temperature simulations with MLFFs to be a consistency check on the values of such parameters.}

\textcolor{black}{In principle, it is possible to go beyond DFT+U and base a MLFF on first-principles calculations with a more elaborate description of the electronic structure, such as, for instance, hybrid (range-separated) functionals   \cite{Silva2007, Tran2006, Heyd2003}, or the random phase approximation (RPA) \cite{Bohm1953, Gell-Mann1957, Langreth1975, Chen2017-rpa}. However, these methods are computationally much more expensive.}

\textcolor{black}{The MD runs in the training phase are not sufficiently accurate to capture the magnetic phase transitions. For instance, the N\'{e}el temperature of LaMnO$_3$ is 140 K, whereas the AFM ordering persists up to $\sim 500$ K. It should be noted, however, that only in the training phase of the force field, where one has access to the DFT electronic structure, the full information on the magnetic ordering is available. The training phase uses a small supercell and it is quite conceivable that this supercell is too small to properly describe a magnetic phase transition.}

\textcolor{black}{Nevertheless, the initial magnetic properties of LaMnO$_3$ and SrRuO$_3$ are important the structures and the structural phase transitions}. At low temperature, LaMnO$_3$ and SrRuO$_3$ have an AFM and FM ordering of magnetic moments on the transition metal ions, respectively, and the MLFFs have been trained starting from these ground states. Starting from an incorrect ground state to train a MLFF, FM ordering in case of LaMnO$_3$, for instance, gives a temperature for the phase transition that is significantly too low, and starting from an unpolarized ground state does not give a phase transition at all, see Supporting Material, Sec. S4. There is an intimate connection between the electronic structure and the magnetic ordering, and a correct electronic structure \textcolor{black}{at low temperature in the training phase} is required for capturing the structural phase transitions \textcolor{black}{in the production phase} correctly.


\section{Conclusion}
Complex oxides are frequently modeled with DFT+U, where the $U$ should capture the physics of the complex interplay between crystal structure, magnetism, and electron {\color{black} localization}. Here we use the new on-the-fly machine learning force field approach to train ML potentials by DFT+U on two archetypal perovskite oxide materials, LaMnO$_3$ and SrRuO$_3$, with transition metals in d$^4$ configuration. For the antiferromagnetic insulator LaMnO$_3$, \textcolor{black}{with the correct value for $U$ and antiferromagnetic ordering}, the model is able to correctly predict the structural phase transition and the suppression of the JT distortion around 750 K. Likewise, for the ferromagnetic metal SrRuO$_3$, the structural phase transitions simulated with the MLFF are in good agreement with experiment. We show that the physics of complex oxides can be captured sufficiently well by DFT+U to predict these phase transitions. \textcolor{black}{Furthermore, we demonstrate how the crystal structure at finite temperatures depend on the parameter $U$. This suggests that by comparing the experimental crystal structure data with simulations, active learning of ML interatomic potentials could serve as an additional approach to assess values for $U$.} 

\textcolor{black}{
\section{Data availability}
Reserach data for this paper has been made available trough a dataset in the 4TU.ResearchData repository, see Ref.~\cite{Jansen2023_database}. The following data is stored: i) The electronic structure databases, including structures and corresponding DFT+U energies, forces and stress tensors, used to train the MLFFs (ML\_AB files). ii) A high level analysis of the electronic structure databases presented by pdf factsheets generated with open-source FPdataViewer software\cite{FPdataViewer23}. iii) VASP input files (INCAR, KPOINTS) corresponding to the on-the-fly MLFF generation.
}

\end{document}